\documentclass[english, pra, twocolumn,superscriptaddress,secnumarabic,aps,nobibnotes,numerical,floatfix]{revtex4-1}

\usepackage{tabularx}
\usepackage{enumerate}
\usepackage[11pt]{moresize}
\usepackage[normalem]{ulem}
\usepackage{amssymb}
\usepackage{xcolor}
\usepackage{amsmath}
\usepackage{float}
\usepackage{amstext}
\usepackage{graphicx}
\usepackage{esint}
\usepackage{bbold}
\usepackage{color}
\usepackage{physics}
\usepackage[colorlinks]{hyperref}
\usepackage{amsthm}
\usepackage{times}
\usepackage{mathrsfs }
\usepackage{bbm}
\usepackage[toc,page]{appendix}
\usepackage{booktabs}
\usepackage{mathtools}
\usepackage[caption=false]{subfig}
\usepackage[colorlinks]{hyperref}

\newcommand{\smx}{\sigma^x}

\newcommand{\smz}{\sigma^z}
\newcommand{\smp}{\sigma^+}
\newcommand{\smm}{\sigma^-}

\newcommand{\half}{\frac{1}{2}}

\newcommand{\hc}{\text{h.c.}}

\newcommand{\Heff}{H_{\rm eff}}

\newcommand{\ophi}{\omega_\Phi}
\newcommand{\wtilde}{\widetilde{\omega}}

\newcommand{\otb}{\omega_{\rm\small c}}

\begin{document}

\title{Analysis of parametrically driven exchange-type (iSWAP) and two-photon (bSWAP) interactions between superconducting qubits}
\author{Marco Roth}
\affiliation{Institute for Quantum Information, RWTH Aachen University, D-52056 Aachen, Germany}
\author{Marc Ganzhorn}
\author{Nikolaj Moll}
\author{Stefan Filipp}
\author{Gian Salis}
\affiliation{IBM Research -- Z\"urich, 8803 R\"uschlikon, Switzerland}
\author{Sebastian Schmidt}
\affiliation{Institute for Theoretical Physics, ETH Z\"urich, 8093 Zurich, Switzerland}

\date{\today}
\begin{abstract}
A current bottleneck for quantum computation is the realization of high-fidelity two-qubit quantum operations between two and more quantum bits in arrays of coupled qubits. Gates based on parametrically driven tunable couplers offer a convenient method to entangle multiple qubits by selectively activating different interaction terms in the effective Hamiltonian. Here, we study theoretically and experimentally a superconducting qubit setup with two transmon qubits connected via a capacitively coupled tunable bus. We develop a time-dependent Schrieffer-Wolff transformation and derive analytic expressions for exchange-interaction gates swapping excitations between the qubits (iSWAP) and for two-photon gates creating and annihilating simultaneous two-qubit excitations (bSWAP). We find that the bSWAP gate is generally slower than the more commonly used iSWAP gate, but features favorable scalability properties with less severe frequency crowding effects, which typically degrade the fidelity in multi-qubit setups. Our theoretical results are backed by experimental measurements as well as exact numerical simulations including the effects of higher transmon levels and dissipation. 
\end{abstract}
\maketitle

Quantum computation is based on accurate and precise control of quantum bits and their interactions to create multi-qubit superpositions and entanglement. With superconducting circuits, single qubit quantum gates can be carried out with fidelities approaching $99.99\%$ \cite{Chow2010,Chow2012,Barends2014,McKay2016_2}, while errors in two-qubit operations are typically higher with record fidelities around 99\% \cite{Barends2014,Sheldon2016}. However, the realization of qubit operations with even higher fidelity is required both for reaching the error threshold for quantum computation \cite{Devoret2013,Corcoles2015,Kelly2015,Riste2015} and for carrying out reliable quantum simulations and optimizations in large arrays of coupled qubits \cite{Schmidt2013,Barends2015,Barends2016,Kandala2017}. Moreover, the quest for useful quantum computations before full quantum error correction becomes available may be assisted by efficient, short-depth gate sequences based on two- or multi-qubit gates \cite{Liu2007, Mezzacapo2014} with versatile types of interactions. In particular, parametric schemes based on tunable couplers have been proposed and recently realized as a means to achieve fast gates with high fidelities \cite{Liu2006a, Bertet2006, Niskanen2007, Tian2008, Allman2010a, Zakka-Bajjani2011, Strand2013, Kapit2013, Allman2014, Sirois2015, McKay2016, Roushan2017, Caldwell2017, Didier2017, Reagor2017, Lu2017}.\\

In this context, effective interactions were engineered in Ref.~\cite{McKay2016} between two transmon qubits mediated by a third, ancilla transmon device (bus), which couples dispersively to both qubits and whose frequency is modulated by an external magnetic flux. Such a flux-modulation scheme provides frequency-selectivity and allows to use fixed-frequency computational qubits, thereby minimizing the sensitvity of the device with respect to magnetic flux noise and disorder effects.
For example, modulating at the (fixed) difference frequency of the qubits brings these qubits effectively into resonance in a co-rotating frame such that a single excitation can be swapped efficiently (iSWAP). The effective Hamiltonian in this case is $H_{\rm{iSWAP}}~\propto {\rm XX + YY}$. 
With this method gate fidelities of $97\%$ have been shown for gates lasting less than $200~\rm{ns}$ \cite{McKay2016}. Unfortunately, for shorter pulses it becomes hard to avoid unwanted transitions to higher-excited qubit levels as well as  qubit-coupler transitions occurring at similar frequencies.
In fact, the transition between the two-qubit excited state $|11\rangle$ and the second-excited state of either qubit ($|02\rangle$ or $|20\rangle$) is separated in frequency only by the anharmonicity of the transmon. Moreover, red-sideband transitions between qubits and coupler may be driven by multi-photon resonances of the drive \cite{Strand2013}. Such frequency-crowding effects motivate the study of gates which operate in different frequency bands sufficiently detuned from those excitations.\\

In this work, we focus on driving the two-photon, blue-sideband transition $|00\rangle\leftrightarrow |11\rangle$ (bSWAP) by modulating the tunable bus at the sum frequency of the qubits \cite{Bertet2006,Niskanen2007,Poletto2012,Filipp2011a}. The effective Hamiltonian in this case is $H_{\rm{bSWAP}}~\propto {\rm XX - YY}$. In contrast to driving at the difference frequency (typically around $1~\rm{GHz}$ or less) for the exchange-interaction gate (iSWAP), the drive is at an elevated frequency around $10~\rm{GHz}$. Higher-harmonics are then pushed to higher frequency ranges without causing spurious drivings. 
A similar gate was realized in Ref.~\cite{Poletto2012} by directly addressing the two-photon transition of the computational qubits via external microwave drives. Here, we demonstrate significantly faster gate times when the bSWAP interaction is mediated parametrically by a flux-modulation of the tunable coupler using a similar device as in Ref.~\cite{McKay2016}.\\

On the theory level, fast modulations at elevated frequencies pose a challenge for methodology as commonly employed adiabatic approximations break down. In order to compare experimental results with theory, we develop a generalized, time-dependent Schrieffer-Wolff transformation \cite{Goldin2000,Theis2016}, which explicitely incorporates the time-dependence of the ancilla transmon frequency as well as counter-rotating terms in the coupling between ancilla and computational transmons \cite{Zhu2013}. Such a transformation eliminates the coupler degree of freedom and yields an effective time-dependent two-qubit Hamiltonian which is valid in a broad range of modulation frequencies. We benchmark the effective model by comparing to exact numerical simulations of the full circuit including higher transmon levels and find excellent agreement.
Analytical estimates for the gate times of iSWAP and bSWAP gates are found in good agreement with numerical simulations as well as experimental measurements.\\

The remainder of the paper is structured as follows. In Section~\ref{sec:model}, we introduce the model of our circuit and its experimental implementation. We also present measurement results for iSWAP and bSWAP gate as well as numerical simulations of the model. In Section \ref{sec:dispersive_regime} we develop the time-dependent Schrieffer-Wolff transformation and provide simple analytic estimates for the effective interaction strengths of both gates. Finally, we compare experimental, numerical and analytical results in Section~\ref{sec:results}, discuss the effects of dissipation and gate errors in Section~\ref{sec:errors} and conclude in Section~\ref{sec:summary}.

\section{Model and Experimental setup}
\label{sec:model}
\begin{figure}[t]
	\includegraphics[width = 0.3\paperwidth]{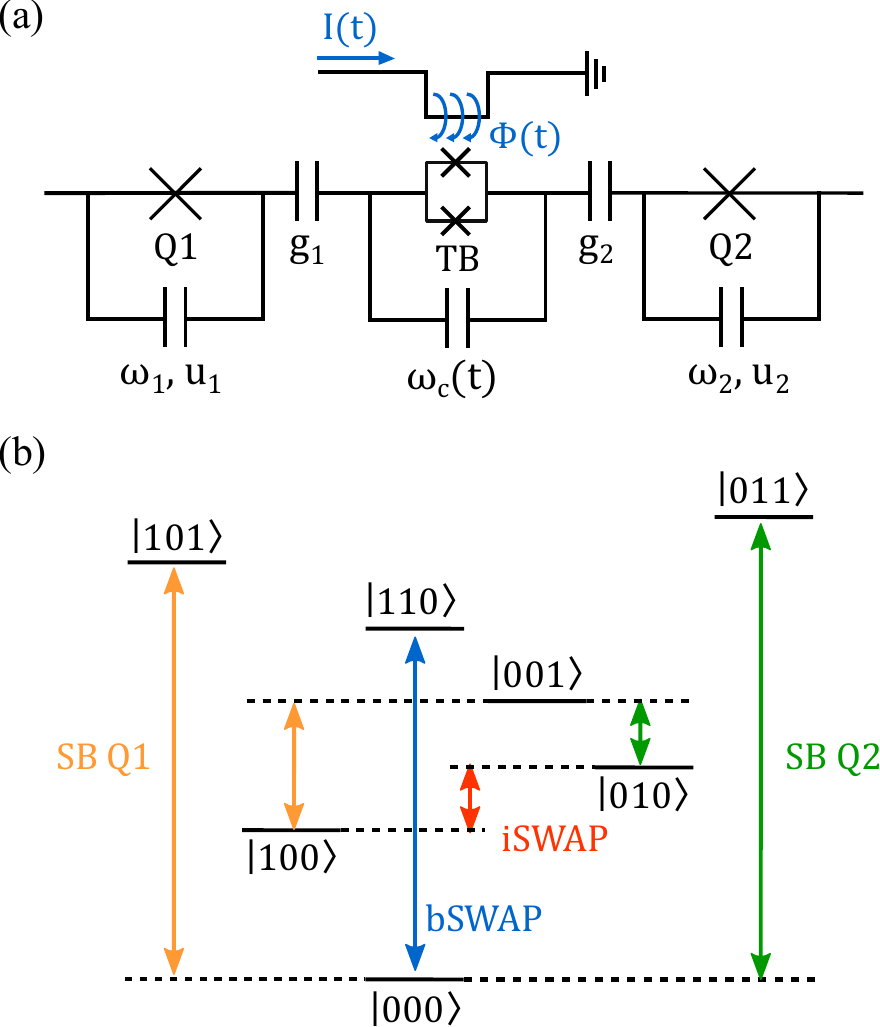}
	\caption{(a) Circuit scheme consisting of two fixed frequency transmons (Q1, Q2) capacitatively coupled to a flux tunable transmon (TB). The tunable coupler is controlled by a high-speed flux line providing a current $I(t)$ and a consequent flux $\Phi(t)$ threading the SQUID-loop of the coupler. Each of the fixed frequency qubits is coupled to an individual readout resonator (not shown). (b) Level diagram of the device. Here, $\ket{n_1 n_2 n_c}$ denotes the state of the combined system with the qubit excitation number $n_{1,2}$ and the coupler excitation number $n_c$. The computational subspace is spanned by the states $\lbrace{\ket{000},\ket{010},\ket{100},\ket{110}\rbrace}$ with the coupler mostly residing in its ground state. The states $\ket{100}$ and $\ket{010}$ are separated by the difference frequency $\sim \omega_1 - \omega_2$ whereas the states $\ket{000}$ and $\ket{110}$ are separated by the sum frequency $\sim \omega_1+\omega_2$. Additionally, the most important sideband transitions between qubits and coupler are shown. }
	\label{fig:setup}
\end{figure}
\begin{table}
\label{fig:table}
\centering
\resizebox{0.47\textwidth}{!}{%

\begin{tabular}{lcccccc}
\hline\hline

   & $\omega/2\pi$ (GHZ) &  $u/2\pi$ (MHz)    &  $g/2\pi$ (MHz) & $T_1$ ($\mu$s) & $T_2$ ($\mu$s) & $T_2^*$ ($\mu$s) \\
Q1 & 4.422     & -349          & 109        & 71             & 52             & n/a \\
Q2 & 4.999     & -330          & 117        & 59             & 32             & n/a \\
TB & 6.006     &  n/k          & n/a        & 11.6           & n/k            & 0.5 \\
\hline\hline
\end{tabular}}
\caption{The device parameter of two fixed-frequency transmons (Q1, Q2) coupled via a flux-tunable transmon (TB) as shown in Fig.~\ref{fig:setup}. The qubits exhibit frequencies $\omega$, anharmonicities $u$, and capacitive couplings $g$. At the flux bias point $\theta= -0.108\, \Phi_0$ the decoherence times $T_1$, $T_2$  and $T_2^*$ indicate that the coherence of the qubits is limited by magnetic flux noise with a power spectral density $S(\omega) = A^2/\omega$, where $A$ is measured to be $A = 3 \cdot10^{-5}\, \Phi_0$. Note, that some parameter values (n/k) could not be measured due to the absence of a readout resonator that directly couples to the tunable bus (TB).}
\end{table}

We study the three-qubit device schematically shown in Fig.~\ref{fig:setup}(a) and described by the Hamiltonian
\begin{equation}
H = -\sum_{i=1,2} \frac{\omega_i}{2} \smz_i -\frac{ \otb(t)}{2} \smz_{\rm\small c} + g_1\smx_1\smx_{\rm c}+ g_2\smx_2\smx_{\rm c}\,,
\label{eq:H_tla}
\end{equation}
where $\sigma^\alpha_i$ denote the standard Pauli operators ($\alpha=x,y,z$). Here, the frequencies $\omega_i$ of the two computational qubits ($i=1,2$) are fixed, 
while the frequency $\otb(t)$ of the coupler qubit in the middle is tunable via an external time-dependent magnetic flux $\Phi(t)$.

\begin{figure*}[t]
	\centering
		\subfloat[]{\includegraphics[width = 0.6\paperwidth]{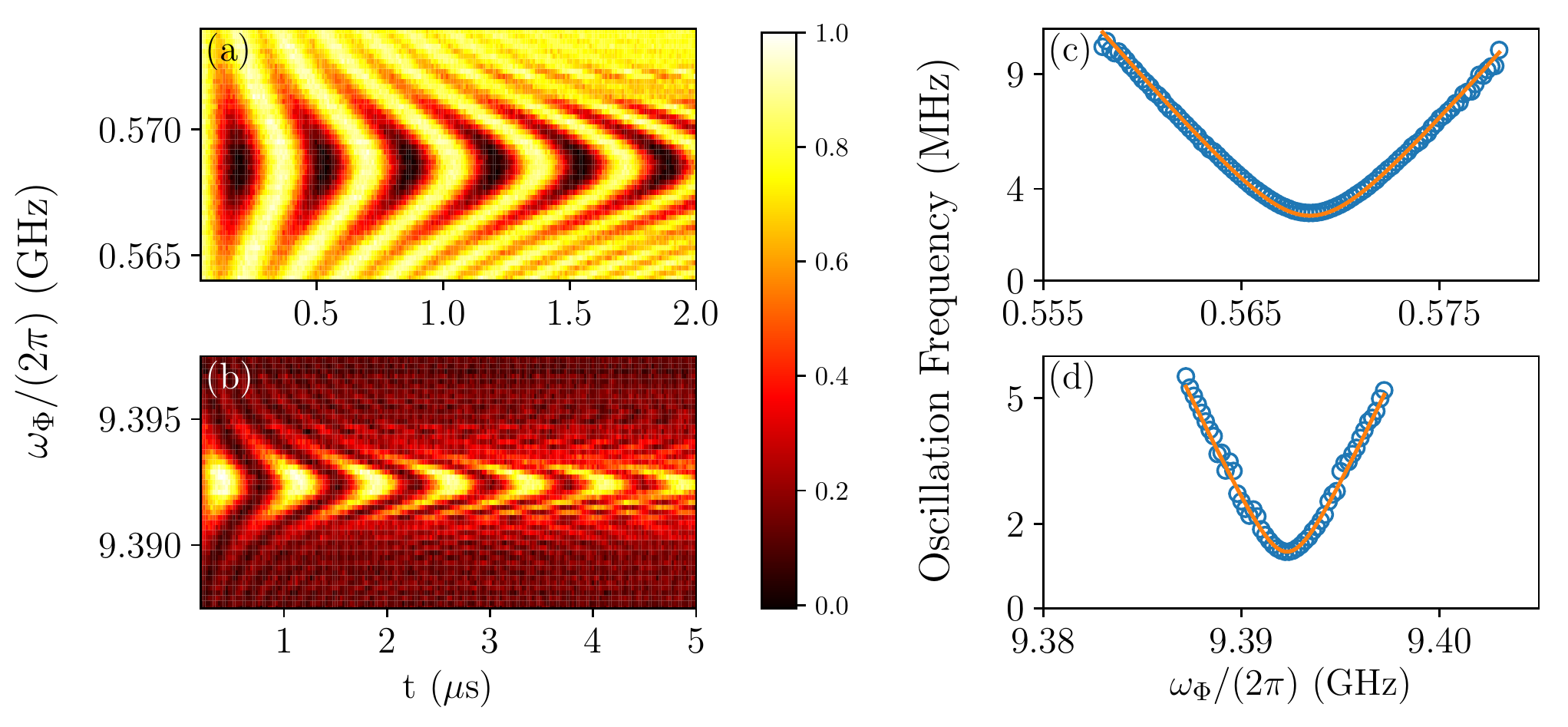}}
	\caption{Left panel: (a) Experimentally measured state population of the $\ket{100}$ state for the iSWAP gate in panel (a) and the $\ket{110}$ state for the bSWAP gate in panel (b) as a function of modulation frequency $\ophi$ and time $t$. The modulation strength is fixed at $\delta = 0.065\,\Phi_0$ in (a) and $\delta = 0.147\,\Phi_0$ in (b). Right panel: Oscillation frequencies of the iSWAP gate in (c) and the bSWAP gate in (d) as obtained from the state occupations shown in (a) and (b), respectively. The solid line is the best fit of the oscillation frequencies (circles) to an effective two-state model.}
	\label{fig:experimental_gates}
\end{figure*}
The flux-dependence of the coupler frequency $\otb(t)$ mediated by an external flux bias line is given by \cite{Koch2007}
\begin{equation}
\otb(t) = \otb^0\sqrt{\abs{\cos(\pi\Phi(t)/\Phi_0)}}\,,
\label{eq:wtb}
\end{equation}
where $\otb^0$ corresponds to the frequency at zero applied flux and $\Phi_0=h/(2e)$ denotes the flux quantum. The computational qubits are both capacitively interacting with the coupler qubit via XX-type interactions with strength $g_i$.
In the following, we consider a harmonic modulation of the external flux, i.e,
\begin{eqnarray}
\Phi(t) = \theta + \delta\cos(\ophi t)
\end{eqnarray}
with the dc bias $\theta$ and an ac component with modulation frequency $\ophi$ and strength $\delta$. For weak modulations ($\delta \ll \theta$), it is useful to expand the coupler frequency $\otb(t)$
in (\ref{eq:wtb}) to second order in the modulation strength $\delta$, i.e., 
\begin{align}
\otb(t)&\approx\omega_c^\theta + \delta \pdv{\otb}{\Phi}\Big|_{\Phi=\theta}\cos(\ophi t) \nonumber\\
&+ \frac{\delta^2}{2} \pdv[2]{\otb}{\Phi}\Big|_{\Phi=\theta}\cos^2(\ophi t) 
\label{eq:otbexpand}
\end{align}
with $\omega_c^\theta=\otb^0\sqrt{\abs{\cos(\pi\theta/\Phi_0)}}$. By a proper choice of the modulation frequency $\ophi$, it is possible to activate various excitations between the computational qubits (e.g., iSWAP, bSWAP and higher transmon levels) as well as between qubits and the coupler (sidebands) as shown in Fig.~\ref{fig:setup}(b). The strength of these excitations is tunable via the dc bias $\theta$ as well as the modulation strength $\delta$.\\

The experimental implementation of model (\ref{eq:H_tla}) consists of two fixed-frequency transmons (Q1, Q2 in Fig.~\ref{fig:setup}) coupled via a flux-tunable transmon (TB in Fig.~\ref{fig:setup}) similar to the setup presented in \cite{McKay2016}. Spectroscopic measurements of our device yield qubit frequencies, capacative coupling strengths and decay rates as summarized in Table~\ref{fig:table}. To perform gate operations we initialize the system in the state $\ket{100}$ for the iSWAP gate and $\ket{000}$ for the bSWAP gate. Experimentally, the $\ket{100}$ state is prepared by applying a $\pi$ pulse to the first qubit. Subsequently, the flux is modulated at a frequency $\ophi$ for a time $T$ with an envelope amplitude consisting of a square pulse with gaussian rise and fall of about $\sim 20\,\rm ns$. Fig.~\ref{fig:experimental_gates}(a,b) show the measured state occupations for this protocol for varying modulation frequencies close to the expected resonances at (a) $\ophi \approx \omega_1-\omega_2$ for the iSWAP and (b) $\ophi \approx \omega_1+\omega_2$ for the bSWAP interaction. Optimal gate performance is achieved at resonance frequencies that are slightly detuned from these simple expressions due to dispersive shifts induced by the coupler, which will be discussed in the next section.
For the modulation strengths $\delta$ chosen in Fig.~\ref{fig:experimental_gates}, we find that the iSWAP interaction is roughly a factor of $2$ faster than the bSWAP gate. The resonance corresponding to the bSWAP interaction is narrower when compared to the iSWAP gate and thus requires a more delicate fine tuning of the modulation frequency.\\
Fig.~\ref{fig:experimental_gates}(c,d) is obtained from fitting the time-dependent data in (a) and (b) with a decaying harmonic oscillation. Shown is the oscillation frequency, i.e., gate strength, as a function of the modulation frequency. The minima of the oscillation frequencies correspond to the highest gate fidelities with gate strength of about $2.8\, {\rm MHz}$ for the iSWAP and about $1.3\, {\rm MHz}$ for the bSWAP gate.

As indicated above, harmonic modulation of the coupler leads not only to the iSWAP and bSWAP transitions as shown in Figure 2, but to a variety of other transitions involving zero or more photons.
To identify the transitions in the vicinity of the wanted iSWAP and bSWAP, we carry out numerical simulations of the model in Eq.~(\ref{eq:H_tla}) including higher transmon levels (see Appendix~\ref{sec:appendix1}).
Fig.~\ref{fig:colorplotFFT}(a,b) shows the oscillation frequencies of the most important resonances in a broader frequency range as compared to Fig.~\ref{fig:experimental_gates}(c,d), e.g., iSWAP, bSWAP, sidebands and higher transmon level excitations. 
In particular, multi-photon sideband transitions involving either qubit 1 or qubit 2 are quite close to the iSWAP gate in Fig.~\ref{fig:colorplotFFT}(a). Most importantly, the oscillation frequencies and associated weights of these sideband transitions are not necessarily small such that a nearby anti-crossing can cause additional leakage errors. This might lead to serious limitations of the iSWAP architecture, when the device is scaled up to couple more than two qubits with different resonance frequencies, e.g., due to disorder effects. By comparing Fig.~\ref{fig:colorplotFFT}(a) with Fig.~\ref{fig:colorplotFFT}(b), it becomes clear that frequency crowding effects turn out less severe for the bSWAP gate. The enhanced free spectral range around the bSWAP gate frequency may thus lead to favorable scaling behaviour.

These initial experimental and numerical results motivate a more in-depth study and comparison between iSWAP and bSWAP gates, in particular of their gate strengths and fidelities.
In the next section, we develop an effective analytic theory, which explains the different magnitude of the interaction strength for iSWAP and bSWAP gate. Later, in Section \ref{sec:errors}, we 
analyze numerically how nearby sidebands and dissipation affect the gate fidelities.

 \begin{figure}[t]
   \includegraphics[width = 0.47\textwidth]{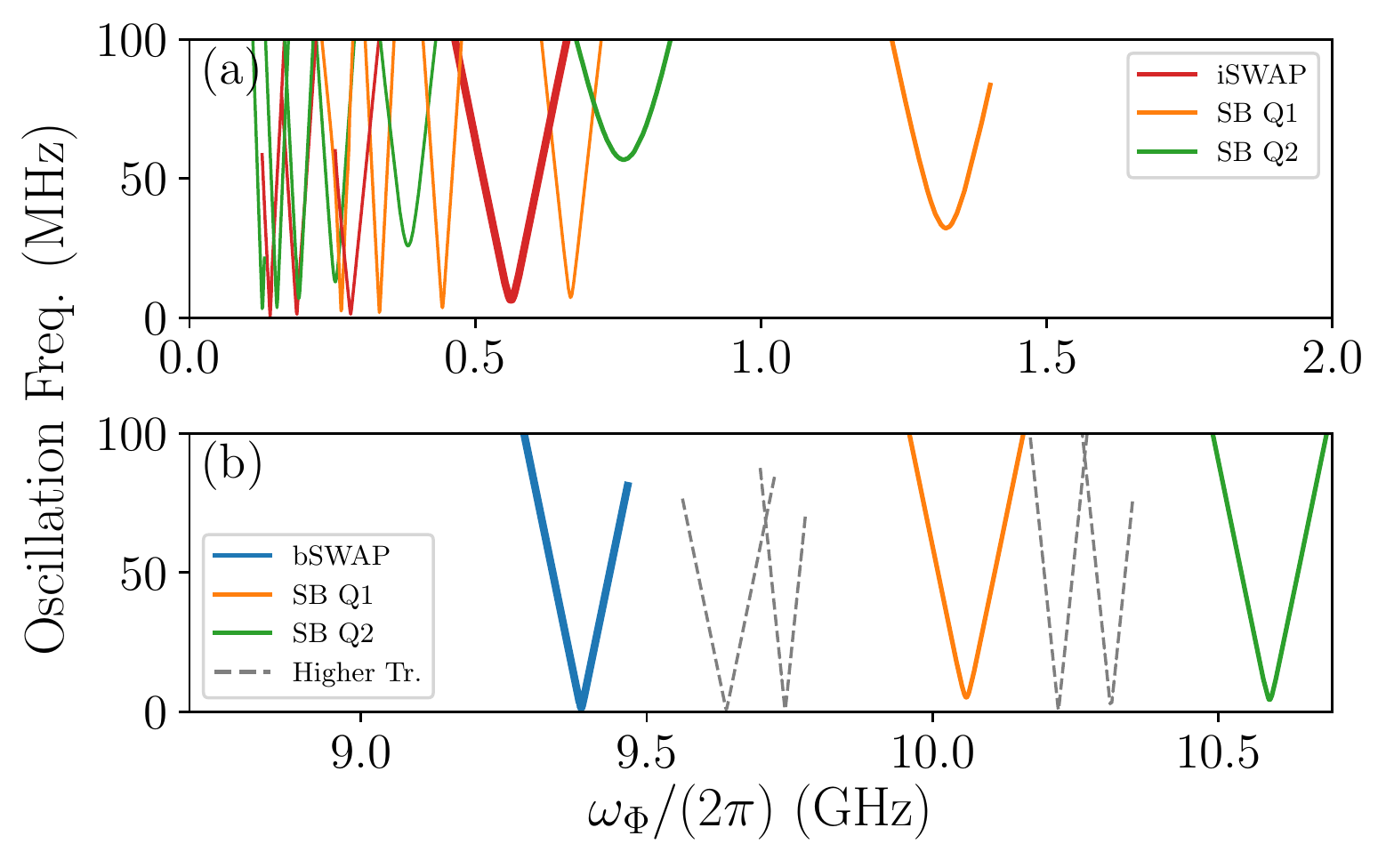}
    \caption{(a) Oscillation frequencies of the population of the $\ket{010}$ state in (a) and the $\ket{000}$ state in (b)  as a function of modulation frequency $\ophi$. The results are obtained by a numerical simulation of the transmon Hamiltonian (\ref{eq:full_transmon_H}) with the coupler frequency approximated by the expansion in Eq.~(\ref{eq:otbexpand}) to second order in $\delta$. The system is initialized in the $\ket{100}$ state in panel (a) and the $\ket{000}$ state in panel (b). Shown are the transitions that result in a population leakage $>10^{-5}$ out of the two-state subspace corresponding to the iSWAP gate in (a) and the bSWAP gate in (b). The DC flux bias is $\theta = -0.108\,\Phi_0$. The flux modulation amplitude is fixed at $\delta = 0.13\,\Phi_0$ in (a) and at $\delta = 0.16\,\Phi_0$ in (b).}
      \label{fig:colorplotFFT}
\end{figure}

\section{Time-dependent Schrieffer-Wolff transformation}
\label{sec:dispersive_regime}
The Schrieffer-Wolff transformation (SWT) is a widely applied perturbative technique in order to eliminate degrees of freedom whose excitations are energetically far detuned
from the Hamiltonian subspace of interest \cite{Bravyi2011}. Time-dependent generalizations were developed in the context of solid-state impurity physics \cite{Goldin2000} and more recently
for the description of parametrically activated quantum gates \cite{Theis2016}. Below, we generalize the previous efforts in order to include excitation non-conserving terms
beyond the rotating-wave approximation \cite{Zhu2013}. The resulting Hamiltonian is valid in a broad frequency range and can describe iSWAP, bSWAP as well as Ising-type interactions.

\subsection{Effective Hamiltonian}

We perform a SWT with $U(t)=\exp{S(t)}$, where the anti-hermitian operator $S^\dagger=-S$ is designed to effectively decouple
the coupler qubit from the rest of the circuit.  For this purpose, the unitary transformation is carried out perturbatively in the dispersive regime,
where qubits are far detuned from the coupler. More specifically, we choose the ansatz
\begin{align}
S(t) =\sum_{i=1,2}\left(\alpha_{i,-}(t)\smp_i \sigma^-_{\rm c} + \alpha_{i,+}(t)\smp_i \sigma^+_{\rm c}- \hc \right)\,,
\label{eq:ansatz_S}
\end{align}
where the parameters $\alpha_{i,\pm}$ fulfill the ordinary differential equation (see Appendix~\ref{sec:appendix2})
\begin{align}
i\dot{\alpha}_{i,\pm} + g_i -  \Delta_{i,\pm}(t) \alpha_{i,\pm}(t) &= 0
\label{eq:deq_alpha}
\end{align}
with $\Delta_{i,\pm}(t)=\omega_i\pm\otb(t)$.
To second order in the coupling strength $g_i$, the effective two-qubit Hamiltonian is given by
\begin{align}
\Heff =&-\sum_{i=1,2} \frac{\wtilde_i(t)}{2}\smz_i \nonumber\\
&+ \left(\Omega_-(t)\smp_1\smm_2  + \Omega_+(t)\smp_1\smp_2 + \hc\right) 
\label{eq:heff_dispersive}
\end{align}
with the dispersively shifted fequencies
\begin{align}
\wtilde_i(t)& = \omega_i + g_i \sum_{\mu=\pm}\Re(\alpha_{i,\mu}(t)) 
\label{eq:dispersive_shifts_qubits}
\end{align}
and the coupling parameters
\begin{align}
\label{eq:J_delta}
\Omega_-(t)  =  (-1/&2) \Big( g_1 \alpha_{2,+}^*(t) + g_2 \alpha_{1,+}(t)  \nonumber\\
		  				 & - g_1 \alpha_{2,-}^*(t) - g_2 \alpha_{1,-}(t) \Big)
\end{align}
and 
\begin{align}
\Omega_+(t) =  (-1/&2) \Big(  g_1 \alpha_{2,+}(t) + g_2 \alpha_{1,+}(t) \nonumber\\
					 & - g_1 \alpha_{2,-}(t) - g_2 \alpha_{1,-}(t) \Big)\,.
\label{eq:J_sigma}
\end{align}
In the interaction picture with respect to the non-interacting part of the Hamiltonian $H_0=-(1/2)\sum_{i=1,2} \wtilde_i(t)\smz_i$ ($H_0$ commutes with itself at different times) we obtain
\begin{align}
\Heff(t) = \Omega_-(t) \smp_1\smm_2 e^{i\varphi_-(t)} + \Omega_+(t) \smp_1\smp_2 e^{i\varphi_+(t)} + \hc \,,
\label{eq:eff_int}
\end{align}
where the phases are given by $\varphi_\pm(t) = \int_{0}^t\dd{t'}\Delta_\pm(t')$ with $\Delta_\pm(t)=\wtilde_1(t) \pm \wtilde_2(t)$.  In general, the time-dependent coupling constants and phases in (\ref{eq:eff_int}) can be calculated efficiently
for arbitrary modulations of the flux from a numerical solution of the differential equations in (\ref{eq:deq_alpha}). For a time-independent Hamiltonian, i.e., constant flux, the solution of (\ref{eq:deq_alpha}) is given by the well known
expression $\alpha_{i,\pm}=g/\Delta_{i,\pm}$ with constant $\Delta_{i,\pm}=\omega_i\pm\omega_c$ yielding $\varphi_\pm(t)=\Delta_\pm t$ and equal coupling constants $\Omega_\pm=(g_1g_2/2)(1/\Delta_{1,-}+1/\Delta_{2,-}-1/\Delta_{1,+}-1/\Delta_{2,+})$. The first term in (\ref{eq:eff_int}) thus rotates at the difference frequency $\Delta_-$ of the two qubits, while the second term rotates at the sum frequency $\Delta_+$. Note, that both phases include the dispersive shifts of the qubits. In the general case, where the qubit frequencies are not equal, both types of interactions  average to zero rather quickly. 
However, the time-dependent coupler frequency will induce weak modulations of the coupling constants as well. By properly choosing the modulation frequency one can compensate for the phases in (\ref{eq:eff_int}) and obtain static interactions. In the next section we derive analytic estimates for the corresponding effective interaction strength.

\subsection{XX$\pm$YY couplings}

For the harmonic time dependence obtained from (\ref{eq:otbexpand}) to first order in $\delta$,  we solve the differential equation (\ref{eq:deq_alpha}) for $\alpha_{i,\pm}(t)$ analytically in Appendix C.  
Assuming that the modulation frequency $\ophi$ is sufficiently detuned from the sideband frequencies $\sim\Delta_{i,\pm}^\theta/n$ with $\Delta_{i,\pm}^\theta=\omega_i\pm\omega_c^\theta$ and integer $n=1,2,..$ (corresponding to $n$-photon resonances), we find
\begin{equation}
\alpha_{i,\pm}(t) \approx \sum_{k = -\infty}^\infty \overline{\alpha}_{i,\pm}(k) e^{ik\ophi t}
\label{eq:alpha_fourier}
\end{equation}
with the coefficients
\begin{equation}
\overline{\alpha}_{i,\pm}(k) = g_i\sum_n \frac{J_{k-n}(\mp\lambda) J_n(\pm\lambda)}{n\ophi + \Delta_{i,\pm}^\theta}\,. 
\label{eq:fourier_coefficients}
\end{equation}
Here, $J_n(x)$ is the $n$-th Bessel function of the first kind and $\lambda = (\delta/\ophi) (\partial_\Phi \otb |_{\Phi=\theta})$ is a small, dimensionless parameter. It follows from (\ref{eq:alpha_fourier}) that
the Fourier transform of the coefficients $\alpha_{i,\pm}(t)$ consists of a series of peaks at integer multiples of the modulation frequency $\ophi$ with weights given by (\ref{eq:fourier_coefficients}). 
Inserting the result (\ref{eq:alpha_fourier}) into (\ref{eq:J_delta}) and  (\ref{eq:J_sigma}) yields analogous expansions for the coupling parameters $\Omega_{\pm}(t)=\sum_k \overline{\Omega}_{\pm}(k)e^{ik\ophi t}$ and for the detunings $\Delta_{\pm}(t)=\sum_k \overline{\Delta}_{\pm}(k)e^{ik\ophi t}$ (which determine the phases $\varphi_\pm(t)$). Neglecting higher harmonics in the effective Hamiltonian (\ref{eq:eff_int}), one then obtains static interactions, if the modulation frequency equals the zero-frequency component of the detuning, i.e., $\ophi\approx\overline{\Delta}_{\pm}(k=0)$. Neglecting dispersive shifts, this simply reduces to $\ophi \approx \omega_1\pm\omega_2$. A quantitative more accurate result including dispersive shifts is given by Eq.~(\ref{eq:dispshift}) in Appendix~\ref{sec:appendix4}.
By choosing the minus sign, i.e., $\ophi \approx \omega_1 - \omega_2$, we obtain the effective iSWAP-type Hamiltonian
 \begin{eqnarray}
\Heff \approx \Omega^{-}_{\rm eff} \left( \smx_1\smx_2 - \sigma^y_1\sigma^y_2\right)\,,
\label{eq:heff_iswap}
\end{eqnarray}
where the interaction strength $\Omega^-_{\rm eff}$ is obtained from the $(k=-1)$ frequency components in (\ref{eq:alpha_fourier}), i.e., 
\begin{eqnarray}
\Omega^{-}_{\rm eff} \approx && \delta \frac{g_1 g_2}{4} \pdv{\omega_c}{\Phi}\Big|_{\Phi=\theta} \left( \frac{1}{\Delta_{1,-}^\theta \Delta_{2,-}^\theta} + \frac{1}{\Delta_{1,+}^\theta\Delta^\theta_{2,+}} \right)
\label{eq:iswap_coupling_strength}
\end{eqnarray}
Similarly, by choosing the plus sign, i.e., $\ophi \approx \omega_1 + \omega_2$, we get the effective bSWAP-type Hamiltonian
 \begin{eqnarray}
\Heff \approx \Omega^{+}_{\rm eff} \left( \smx_1\smx_2 + \sigma^y_1\sigma^y_2\right)\,,
\label{eq:heff_bswap}
\end{eqnarray}
with
\begin{eqnarray}
\Omega^{+}_{\rm eff} \approx && -\delta \frac{g_1 g_2}{4} \pdv{\omega_c}{\Phi}\Big|_{\Phi=\theta} \left( \frac{1}{\Delta_{1,-}^\theta \Delta_{2,+}^\theta} + \frac{1}{\Delta_{1,+}^\theta\Delta^\theta_{2,-}}\right)
\label{eq:bswap_coupling_strength}
\end{eqnarray}

We note that the results in (\ref{eq:iswap_coupling_strength})-(\ref{eq:bswap_coupling_strength}) were derived by keeping Besselfunctions of order 0 and 1 in (\ref{eq:fourier_coefficients}) and using the asymptotic expansions $J_0(x)\approx 1$ and $J_1(x)\approx x/2$ for $x\ll 1$, i.e., small modulation strength $\delta$. Consequently, the gate strength are to leading order linear in the modulation strength $\delta$ and directly proportional to the curvature of the frequency-flux relationship $\omega_c \leftrightarrow \Phi$, which is tunable via the dc bias $\theta$. The difference between iSWAP and bSWAP interaction strengths show up inside the brackets in (\ref{eq:iswap_coupling_strength}) and (\ref{eq:bswap_coupling_strength}) and will be discussed in more detail in Section~\ref{sec:results}.

The effective Hamiltonian (\ref{eq:heff_iswap}) corresponds to a rotation in the computational subspace of the qubits only spanned by $\lbrace\ket{01}, \ket{10}\rbrace$ and therefore allows for excitation transfer from one qubit to the other. The effective Hamiltonian (\ref{eq:heff_bswap}) corresponds to a rotation in the subspace spanned by $\lbrace\ket{00}, \ket{11}\rbrace$, a two photon process that would be forbidden in the absence of the coupler. The angle of rotation is given by $\vartheta = \Omega_{\rm eff}^\pm\int_0^T\dd{t'} f(t')$, where the dimensionless parameter $f(t)$ encodes the pulse shape of the flux modulation and $T$ is the length of the flux pulse. The choice $\vartheta = \pi/2$ implements the iSWAP gate for (\ref{eq:heff_iswap}) and the bSWAP gate for (\ref{eq:heff_bswap}). Both gates are Clifford operations and together with single qubit rotations they each form a universal set for quantum computation. We also note, that a modulation scheme, which involves two frequencies,
allows to realize almost arbitrary combinations of $XX$-type and $YY$-type interactions, e.g., Ising gates. The experimental realization of such a complex scheme is left for future work.

\subsection{Adiabatic approximation}
\label{sec:adiabatic_approximation}
It is instructive to compare the results of the previous section with those obtained from a standard adiabatic approximation as employed in \cite{McKay2016}.
In particular, for weak modulations one may be tempted to neglect the time derivative term in (\ref{eq:deq_alpha}) even in the case of a time-dependent coupler frequency
and approximate $\alpha_{i,\pm}(t)\approx g_i/\Delta_{i,\pm}(t)$. Plugging this result back into (\ref{eq:eff_int}) and choosing the modulation frequency as previously, i.e., $\ophi\approx\omega_1\pm\omega_2$,
leads to equal gate strength for iSWAP and bSWAP gates given by

\begin{equation}
\Omega_{\rm ad}^\pm = \delta\frac{g_1g_2}{8}\pdv{\otb}{\phi}\Big|_{\phi=\theta}\sum_{i=1,2}\left(\frac{1}{(\Delta^\theta_{i,-})^2}+\frac{1}{(\Delta^\theta_{i,+})^2}\right)\,.
\label{eq:adiabatic_coupling}
\end{equation}
A condition for the validity of this result is obtained by substituting $\alpha_{i,\pm}(t)\approx g_i/\Delta_{i,\pm}(t)$ back into (\ref{eq:deq_alpha}). The derivative term is then negligible as long as
 $\partial \omega_{c}/\partial t \ll \Delta_{i,\pm}(t)^2$. Together with (\ref{eq:otbexpand}) this yields
\begin{equation}
\abs{\delta \pdv{\otb}{\Phi} }_{\Phi=\theta} \ophi \ll \abs{\Delta_{i,\pm}^\theta}^2\,.
\label{eq:adiabatic_condition}
\end{equation}
Consequently, an adiabatic approximation requires not only small modulation strength $\delta$ but also sufficiently small modulation frequency $\ophi$. In particular, the latter requirement is hardly fulfilled in the case of a bSWAP interaction. 
We note that the adiabatic result in (\ref{eq:adiabatic_coupling}) can also be recovered from (\ref{eq:fourier_coefficients}) by first taking the limit of $\delta\rightarrow 0$ at fixed $\ophi$ when expanding the Bessel functions and afterwards taking the limit of vanishing modulation frequency $\ophi\rightarrow 0$. 
\begin{figure}[t]
	\centering
	\subfloat[]{\includegraphics[width = 0.4\paperwidth]{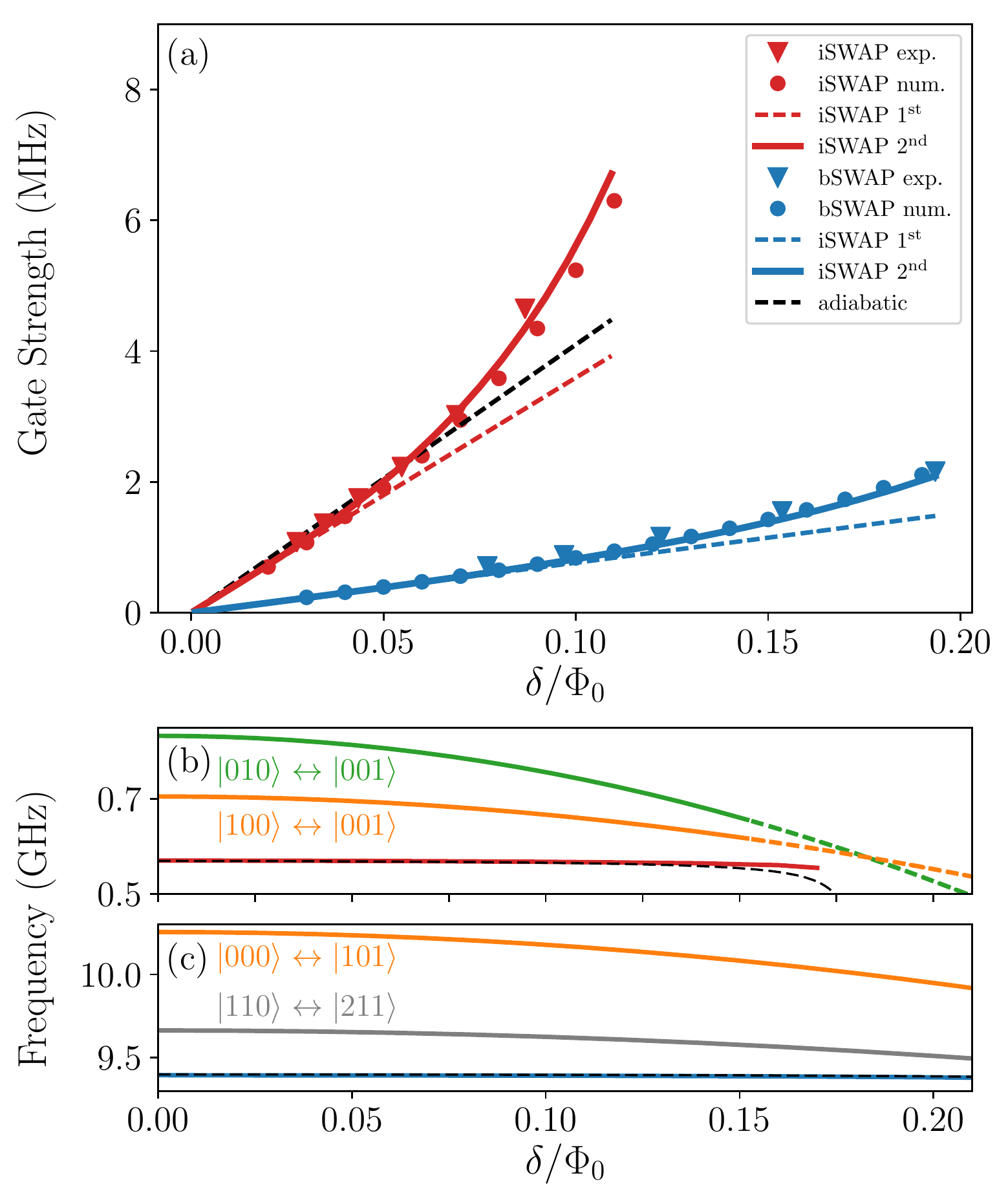}	}
	\caption{(a) Gate strength of iSWAP (red) and bSWAP gate (blue) as a function of $\delta/\Phi_0$. Experimental data is shown as triangles. Circles show the numerical simulation of the full transmon Hamiltonian (\ref{eq:full_transmon_H}). In contrast to Fig.~\ref{fig:colorplotFFT}, the numerical simulations have been carried out using the complete transfer function in Eq.~(\ref{eq:wtb}). Dashed lines show the analytic results to linear order in $\delta$, i.e., Eq.~(\ref{eq:iswap_coupling_strength}) and Eq.~(\ref{eq:bswap_coupling_strength}). Solid lines show analytic results up to second order in $\delta$ (cf. Appendix \ref{sec:appendix3}). The black-dashed line corresponds to the adiabatic approximation in (\ref{eq:adiabatic_coupling}). The values of $\delta$ for the experimental data points are calibrated using a fit between the measured dispersive shifts of the modulation frequencies and the prediction of the effective model (cf. Appendix \ref{sec:appendix4}). (b) Modulation frequency $\ophi$ for the iSWAP gate obtained from numerical simulations (red) and the effective model in Appendix \ref{sec:appendix4} (black dashed). Additionally, the two closest side-band frequencies are shown (orange,green). For the latter, the coupler frequency $\otb(t)$ is expanded to second order in $\delta$. The overbar indicates a time average, replacing $\cos^2(\ophi t)\rightarrow1/2$ in Eq. (\ref{eq:otbexpand}). (c) shows analogous results for the bSWAP gate (blue) including a nearby sideband (orange) and the excitation of a higher-transmon level (grey).} 
\label{fig:coupling_strengths}
\end{figure}

\section{Comparison of results}
\label{sec:results}
In Fig.~\ref{fig:coupling_strengths}(a) we compare the predictions of the SWT in (\ref{eq:iswap_coupling_strength}) and (\ref{eq:bswap_coupling_strength}) with exact numerical calculations as well as with experimental data. 
For the numerical calculations, we determine the gate strength from the time dynamics of an initial state followed by a modulation pulse as described in Section \ref{sec:model}. 
We repeat these simulations for varying modulation frequencies and extract the maximal gate time corresponding to the highest gate fidelity. 
We find that the effective model yields the correct linear asymptotic of the full numerical simulation for small modulation strength $\delta$ for iSWAP as well as bSWAP interaction.
The iSWAP gate strength in (\ref{eq:iswap_coupling_strength}) is inversely proportional to the product $\Delta_{1,-}\Delta_{2,-}$ which is much smaller than the cross-factors $\Delta_{1,-}\Delta_{2,+}$ and $\Delta_{1,+}\Delta_{2,-}$ appearing in the numerator of the bSWAP gate strength in (\ref{eq:bswap_coupling_strength}) in agreement with the numerical findings. On the contrary, the standard adiabatic approximation fails to predict the correct bSWAP interaction strength, which is roughly a factor of four smaller than the iSWAP interaction when comparing results for the same modulation strength, e.g., $\delta \sim 0.1\Phi_0$.

However, by increasing the modulation strength further, we have measured bSWAP interaction strength of up to $2\, {\rm MHz}$, which is substantially larger than those measured with more conventional, non-parametric schemes \cite{Poletto2012}. To obtain experimental results we have recorded the oscillation frequencies shown in Fig.~\ref{fig:experimental_gates} for different amplitudes of the modulation signal. 
The measured gate times generally compare very well with the numerical simulations and our effective model, especially at small modulation strength.
The remaining discrepancy can be attributed to small errors in the calibration of the $\delta$ scale, which is obtained by fitting the slope of the measured dispersive shifts to the analytical model (Appendix~\ref{sec:appendix4}).
Residual frequency shifts not included in the model, e.g., due to a dispersive coupling to states outside the computational subspace, are accounted for by allowing for a small, constant frequency offset in the fitting procedure.

The two lower panels in Fig.~\ref{fig:coupling_strengths} show the dependence of the resonance frequency on the modulation strength $\delta$ for the iSWAP gate (b) and for the bSWAP gate (c) including two nearby excitations in each case as obtained from analytical and numerical results. With increasing $\delta$ both excitations move closer to the iSWAP and bSWAP gates, which can lead to stronger leakage errors. The effect of nearby sidebands is more severe for the iSWAP gate in the considered parameter regime. 
From Eq. (\ref{eq:fourier_coefficients}) we also see that the Fourier coefficients which enter the gate strength diverge as soon as the sideband excitations get close to the modulation frequency. While the divergence itself is an artifact of the approximation (the perturbative SWT becomes invalid when the Fourier weights become of order unity), it still explains the higher gate strength of the iSWAP gate compared to the bSWAP gate. On the other hand, by choosing the device parameters such that the sum-frequency of the qubits gets closer to the sideband frequencies, one could further increase the interaction strength for the bSWAP gate (at the expense of additional leakage errors). Consequently, there is a trade-off between maximal gate strength and maximal fidelity. This will be analyzed in more detail in the next section. 

\section{Dissipation effects and gate error}
\label{sec:errors}

In this section we discuss error sources, which typically limit the fidelity of iSWAP and bSWAP gates.
The gate error $\epsilon$ is calculated as $\epsilon=1-F$, where $F$ denotes the gate fidelity defined as 
\begin{equation}
F = \int\dd{\psi}\bra{\psi}U_{\rm gate}^\dag\rho U_{\rm gate}\ket{\psi}\,.
\label{eq:gate_fidelity}
\end{equation}
Here, $\dd{\psi}$ is the Haar measure over the computational state space and $U_{\rm gate}$ is the unitary corresponding to an ideal gate operation \cite{Nielsen2002}. The density matrix $\rho$ is obtained from a simulation of 
the Lindblad master equation
\begin{equation}
\label{eq:master}
\dot\rho=-\frac{i}{\hbar}[H_{\rm Tr},\rho]+\sum_{i = 1,2, \rm c} \left[\Gamma_{i}^-\mathcal{L}[a_i]\rho + \Gamma_{i}^z\mathcal{L}[a^\dag_i a_i]\rho\right]
\end{equation}
with the standard Lindblad operator $\mathcal{L}[C]=(2 C \rho(t) C^{\dag} -\acomm{\rho(t), C^{\dag}C}\rho)/2$. Note, that the density matrix used in (\ref{eq:gate_fidelity}) is obtained from the solution of (\ref{eq:master}) by tracing over all 
degrees of freedom except the computational subspace. The decay rates for the i-th transmon are given by the dissipation rates reported in Section~\ref{sec:model} via $\Gamma_{i}^z=(1/2)(1/T_{2,i}-1/(2T_{1,i}))$ and $\Gamma_{i}^-=1/T_{1,i}$.\\
\begin{figure}[t]
	\centering
	\subfloat[]{\includegraphics[width = 0.38\paperwidth]{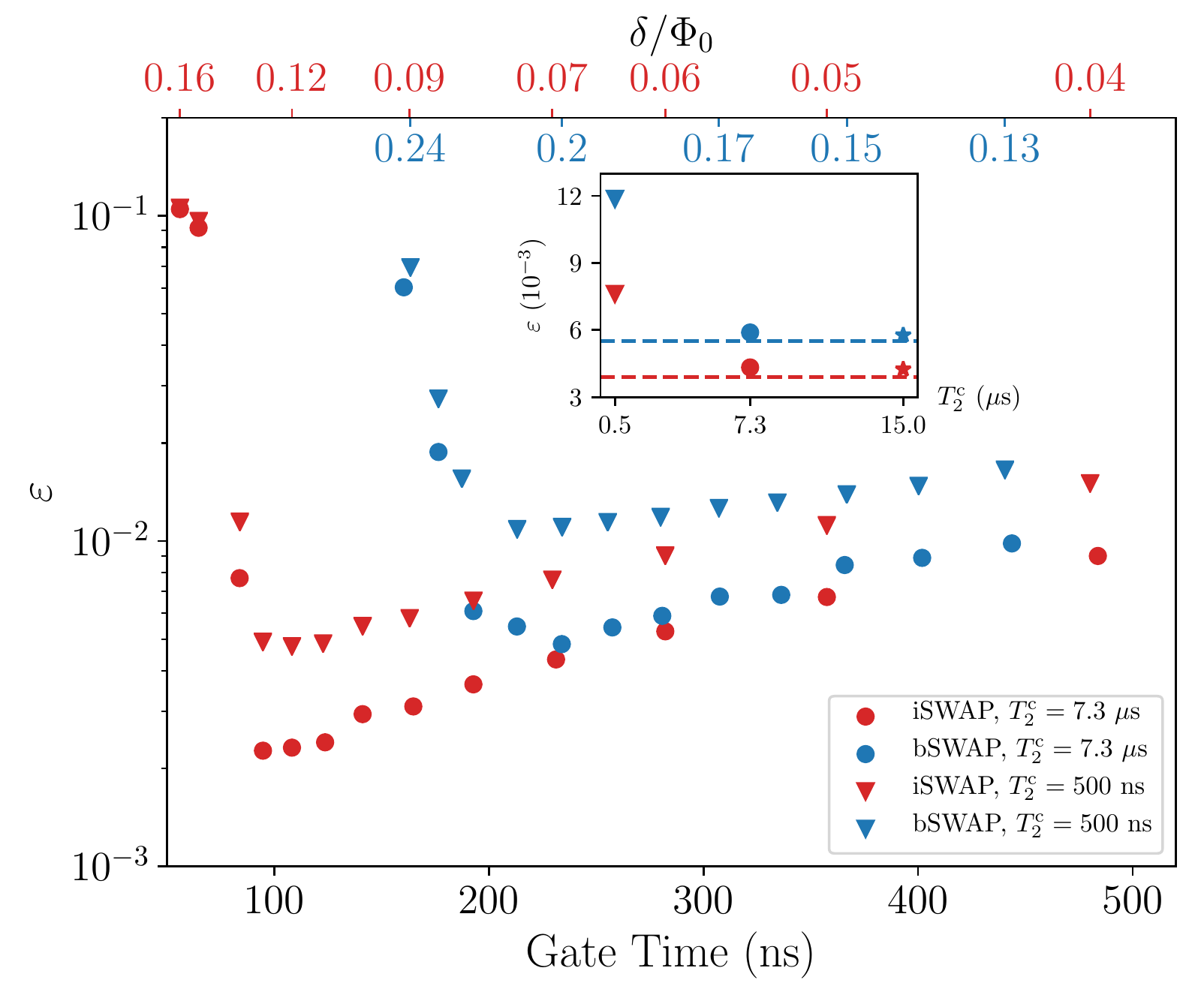}}
	\caption{(a) Error $\epsilon=1-F$ of the iSWAP (red) and bSWAP (blue) gate as a function of gate time. The gate fidelity $F$ is calculated according to Eq.~(\ref{eq:gate_fidelity}) from numerical simulations of the master equation (\ref{eq:master}). The circles in the main figure are obtained with a $T_2$ time of the coupler of $T_2^{\rm c} = 7.3\,\rm \mu s$ whereas the coupler $T_2$ time for the triangles is $T_2^{\rm c} = 500\,\rm ns$. The $\delta$ scale is valid for the triangles. The inset shows the dependence of the error on the $T_2$ time of the coupler for a fixed gate time of $230 {\rm ns}$ (iSWAP) and $280 {\rm ns}$ (bSWAP). The dashed lines in the inset correspond to the results with a dissipation-free coupler element.} 
\label{fig:gate_time_error}
\end{figure}
We simulated the fidelity of both iSWAP and bSWAP gate for two different coupler decoherence times $T_2^c = 7.3 {\rm \mu s}$ and $T_2^c = 0.5 {\rm \mu s}$ to characterize its effect on the gate fidelity.
The results of the simulations are summarized in Fig.~\ref{fig:gate_time_error}. In general, we find two main error sources in agreement with the findings in Ref.~\cite{McKay2016}: (i) decoherence due to relaxation and dephasing and (ii) leakage either into higher transmon
levels or excitations of the coupler (sidebands). For long gate times, i.e., small modulation strength $\delta$, the gate error is mainly limited by the decoherence of the computational qubits. Decoherence caused by the coupler has a comparably weak effect
even for $T_2$ times which are an order of magnitude smaller than those of the computational qubits (see inset of Fig.~\ref{fig:gate_time_error}). 
Only when the $T_2$ time of the coupler becomes comparable to the gate time ($\sim 500 {\rm ns}$), we see a substantial increase of the gate error.
The average leakage population for the longest simulated gate times in Fig.~\ref{fig:gate_time_error} is well below $10^{-6}$. This changes drastically when decreasing the gate time, i.e., increasing the modulation strength $\delta$.
For gate times of about $\sim 100 (200) {\rm ns}$, we observe a sharp increase of the leakage population for the iSWAP (bSWAP) gate. This is consistent with the results shown in Fig.~\ref{fig:coupling_strengths}(b,c), i.e., the closing of the gap between the modulation frequency of the gate
and the resonance frequency of the closest residual excitation. Interestingly for gate times $>~200 {\rm ns}$, the error rates of iSWAP and bSWAP gates are similar even though the bSWAP gate requires a much larger modulation amplitude $\delta$. This can be attributed to the enhanced free spectral range and weaker effects of nearby anti-crossings at elevated modulation frequencies consistent with the results shown in Fig.~\ref{fig:colorplotFFT}. We also note, that no residual excitations exist below the bSWAP in Fig.~\ref{fig:colorplotFFT}(b), corresponding to a total free spectral range of about $\sim 1 {\rm GHz}$
around the bSWAP modulation frequency.\\

\section{Summary \& Outlook}
\label{sec:summary}
In summary, we have studied theoretically and experimentally a parametrically driven two-photon gate (bSWAP) and compared its performance to
the more commonly employed exchange-interaction gate (iSWAP). We find that the bSWAP gate is generally slower than the iSWAP gate at equal parametric amplitudes, but still faster than
realizations based on non-parametric drive schemes. We derive simple analytic expressions for the strength of both gates and find good agreement with numerical
and experimental data. Our calculations suggest that frequency-crowding effects become less severe at elevated frequencies corresponding to the bSWAP gate. 
For the future, we plan to scale-up the current device and investigate  the scalibitliy of the modulation scheme when the coupler mediates two-qubit interactions between several qubits simultaneously.
The combination of iSWAP (XX+YY) and bSWAP (XX-YY) Hamiltonian terms at simultaneous driving also opens up the possibility to engineer Ising-type interactions. 
On the theory level, it would be worthwhile to apply the time-dependent SWT on the density matrix level
in order to obtain an effective two-qubit Liouvillian, which includes dissipative effects and allows for analytic estimates of the gate fidelities.

\begin{acknowledgments}
We thank the quantum team at IBM T. J. Watson Research Center, Yorktown Heights, in particular David McKay for insightful discussions and the provision of qubit devices, and 
David P. DiVincenzo for helpful advice. This work was supported by the IARPA LogiQ program under contract W911NF-16-1-0114-FE and the ARO under contract
W911NF-14-1-0124.
\end{acknowledgments}

\appendix
\section{Higher Transmon Levels}
\label{sec:appendix1}
In Eq.~(\ref{eq:H_tla}) we have modelled the superconducting qubits as two-level systems. In the case of transmon qubits this assumption is often too simplistic and more levels need to be included
for a quantitative accurate description. For the numerical simulations performed in this paper we have used the generalized Hamiltonian
\begin{align}
H_{\rm Tr} &= \sum_{i=1,2} \left[\omega_i a^\dag_ia_i - \frac{u_i}{2}a^\dag_ia_i(a^\dag_ia_i-1)\right]\nonumber\\ 
& + \omega_{\rm c}(t) a^\dag_{\rm c}a_{\rm c} - \frac{u_{\rm c}}{2}a^\dag_{\rm c}a_{\rm c}(a^\dag_{\rm c}a_{\rm c}-1)\nonumber\\
& + \sum_{i=1,2} g_i(a_i^\dag + a_i)(a_{\rm c}^\dag + a_{\rm c})\,,\label{eq:full_transmon_H}
\end{align}
where the creation (annihilation) operators for transmon $i$ are denoted by the bosonic operators $a^\dag_i$ ($a_i$). The anharmonicity of the transmon leads to a Kerr nonlinearity with strength $u_i$ \cite{Koch2007}. By restricting the Hilbert space to the lowest two states of each transmon we obtain Eq.~(\ref{eq:H_tla}).\\
\section{Schrieffer-Wolff transformation}
\label{sec:appendix2}
We consider a unitary transformation of the Hamiltonian $H=H_0+V$ with $U(t)=\exp{S(t)}$ and $S^\dagger=-S$ . 
Assuming that $V$ and $S$ are proportional to a small parameter ($\sim g$), we expand the Hamiltonian in the
new frame $\overline{H}$ to second order in $g$, i.e.,
\begin{align}
\overline{H}&= UHU^\dag - iU\left(\frac{\partial U}{\partial t}\right)^\dag \approx \Heff + H_{\rm V} 
\label{eq:S_expansion}
\end{align}
with
\begin{equation}
\Heff = H_0 + \comm{S}{V} + \half \comm{S}{\comm{S}{H_0}} + \frac{i}{2}\comm{S}{\pdv{S}{t}}\,,
\label{eq:heff_commutators}
\end{equation}
and
\begin{equation}
H_{\rm V} = i\pdv{S}{t}+ \comm{S}{H_0} + V\,.
\end{equation} 
A straightforward calculation shows that the effective Hamiltonian $\Heff$ for $H$ and $S$ given by (\ref{eq:H_tla}) and (\ref{eq:ansatz_S}), respectively, reads 
\begin{align}
\Heff&=\sum_{i=1,2}  -\frac{\wtilde_i(t)}{2}\smz_i - \frac{\wtilde_{\rm c}(t)}{2}\smz_{\rm c}  \nonumber\\
&+ \left(\Omega_-(t)\smp_1\smm_2 + \hc\right)\smz_{\rm c} + \left(\Omega_+(t)\smp_1\smp_2 +\hc\right)\smz_{\rm c}\,.
\label{eq:heff_dispersive}
\end{align}
The Hamiltonian in (\ref{eq:heff_dispersive}) is blockdiagonal, i.e., the Hilbert space of the computational qubits is decoupled from the tunable coupler. Setting $\smz_{\rm c}\approx 1$ and omitting constant term gives the two-qubit Hamiltonian in (\ref{eq:eff_int}). The necessary condition $H_{\rm V} = 0$ then leads to the differential equation (\ref{eq:deq_alpha}). 
\section{Solution of Equation (\ref{eq:deq_alpha})}
\label{sec:appendix3}
The differential equation (\ref{eq:deq_alpha}) can be cast into the general form
\begin{equation}
\dv{y}{t} + P(t)y = Q(t)\,,
\label{eq:general_deq}
\end{equation}
with $y(t)=\alpha_{i,\pm}$, $Q(t) = ig_i$ and $P(t)=i\Delta_{i,\pm}(t)$.
The general solution of (\ref{eq:general_deq}) is given by
\begin{equation}
y=u(t)^{-1}\left[\int u(t')Q(t')\,{dt' }+C_\pm\right]\,,
\label{eq:general_deq:solution}
\end{equation}
with the integration constant $C_\pm$ and the integrating factor $u(t) = \exp(\int P(t')\dd{t'})$.
For the harmonic expansion in (\ref{eq:otbexpand}) we can readily calculate the integrating factor to leading order in $\delta$, i.e., %
\begin{equation}
u(t) = e^{i \Delta_{i,\pm}^\theta t} e^{- i \lambda\sin(\ophi t)}\,,
\label{eq:expression_for_u}
\end{equation}
with $\lambda = (\delta/\ophi) (\partial_\Phi \otb |_{\Phi=\theta})$.
In order to evaluate the remaining integral in (\ref{eq:general_deq:solution}), we utilize the Jacobi-Anger expansion 
\begin{equation}
e^{{\pm i\lambda\sin \theta }}=\sum _{{n=-\infty }}^{{\infty }}J_{n}(\pm\lambda)\,e^{{in\theta }}\,,
\label{eq:jacobi_anger}
\end{equation}
where $J_n(z)$ is the $n$-th Bessel function of the first kind. Plugging (\ref{eq:expression_for_u}) and  (\ref{eq:jacobi_anger}) into (\ref{eq:general_deq:solution}) we obtain
\begin{align}
\alpha_{i,\pm}(t) &= g_i\sum_{m=-\infty}^{\infty}\sum_{n=-\infty}^{\infty}J_m(\mp\lambda) J_n(\pm\lambda)\nonumber\\
&\cross \frac{e^{i(n+m)\ophi t}}{n\ophi +\Delta_{i,\pm}^\theta} + C_{i,\pm}(t)\,.
\label{eq:analytic_alpha_xy}
\end{align}
The initial condition $\alpha_{i,\pm}(0) = g_i/\Delta_{i,\pm}$ yields 
\begin{align}
C_{i,\pm}(t) &=g_i \sum_{m=-\infty}^{\infty} e^{i(m\ophi-\Delta_{i,\pm}^\theta) t}\nonumber\\ &\cross J_m(\mp\lambda)\left(\frac{1}{\Delta_{i,\pm}^\theta}-\sum_{n=-\infty}^{\infty}\frac{J_n(\pm\lambda)}{n\ophi + \Delta_{i,\pm}^\theta }\right)\,.
\label{eq:deq_constant}
\end{align}
As long as the modulation frequencies are sufficiently detuned from the sidebands at $\sim \Delta_{i,\pm}^\theta/n$, the integration constant $C_{i,\pm}(t)$ rotates rapidly and can therefore be neglected in Eq.~(\ref{eq:analytic_alpha_xy}), yielding the approximate result in Eq.~(\ref{eq:alpha_fourier}) and Eq.~(\ref{eq:fourier_coefficients}).\\
We have also included higher order terms in the expansion of the coupler frequency in (\ref{eq:otbexpand}) and performed calculations completely analogous to the steps outlined above in order to obtain analytic solutions of $\alpha_{i,\pm}(t)$ to a higher precision. 
We find that keeping terms up to second order in $\delta$ is sufficient to obtain expressions for $\Omega^\pm_{\rm eff}$ that are almost indistinguishable from the numerical results in Fig.~\ref{fig:coupling_strengths}. The resulting algebraic expressions, however, are lengthy and not particularly insightful beyond 
of what has been stated above. We have therefore omitted them here for brevity.

\section{Dispersive shifts}
\label{sec:appendix4}
By expanding the shifted qubit frequencies  (\ref{eq:dispersive_shifts_qubits}) in a Fourier series $\tilde{\omega}_{i}(t)=\omega_i+\sum_k \overline{\delta\omega}_{i}(k)e^{ik\ophi t}$ (see discussion below Eq.~(\ref{eq:fourier_coefficients})), 
and expanding (\ref{eq:dispersive_shifts_qubits}) self-consistently to second order in $\delta$ yields for the $k=0$ components 
\begin{eqnarray}
\label{eq:dispshift}
\overline{\delta\omega}_i(0) &&\approx  g_i^2\left(\frac{1}{\Delta_{i,-}^\theta}+\frac{1}{\Delta_{i,+}^\theta}\right)-\left(\pdv{\omega_c}{\Phi}\Big|_{\Phi=\theta}\right)^2\frac{\delta^2}{\omega_\pm^2}
\nonumber\\ &&\cross\frac{g_i^2\omega_i\left(\omega_\pm^2-\Delta^\theta_{i,-}\Delta_{i,+}^\theta\right)}{(\omega_\pm^2-(\Delta_{i,-}^\theta)^2)(\omega_\pm^2-(\Delta_{i,+}^\theta)^2)}\,
\end{eqnarray}
with $\omega_\pm=\omega_1\pm\omega_2$. Consequently, the dispersive shifts for the optimal modulation frequencies $\ophi= \omega_1\pm\omega_2+\delta\ophi$ are given by $\delta\ophi\approx\overline{\delta\omega}_1(0)\pm\overline{\delta\omega}_2(0)$. Note that Eq. (\ref{eq:dispshift}) contains the usual Lamb-type shift induced by the tunable bus (first term on the {\rm r.h.s} of Eq.~(\ref{eq:dispshift})) as well as an AC shift induced by the modulation (second term on the {\rm r.h.s} of Eq.~(\ref{eq:dispshift})).

\end{document}